\pdfoutput=1

\documentclass[11pt]{article}

\usepackage[final]{acl}
\usepackage{fancyvrb}
\usepackage{times}
\usepackage{latexsym}

\usepackage[T1]{fontenc}

\usepackage[utf8]{inputenc}

\usepackage{microtype}

\usepackage{inconsolata}

\usepackage{graphicx}

\usepackage[ruled,vlined]{algorithm2e}

\usepackage{longtable}
\usepackage[utf8]{inputenc} 
\usepackage[T1]{fontenc}    
\usepackage{url}            
\usepackage{booktabs}       
\usepackage{amsfonts}       
\usepackage{nicefrac}       
\usepackage{microtype}      
\usepackage{bm}
\usepackage{enumitem}
\usepackage{graphicx}
\usepackage{amsmath}
\usepackage{subfigure}
\usepackage{amssymb}
\usepackage{caption}
\usepackage[most]{tcolorbox}

\DeclareMathOperator*{\argmax}{arg\,max}

\usepackage{amsthm}
\newtheoremstyle{mydefinitionstyle}
  {} 
  {} 
  {\itshape} 
  {} 
  {\itshape} 
  {.}         
  { }         
  {}          

\theoremstyle{mydefinitionstyle}

\newtheorem{definition}{Definition}

\usepackage{array}
\usepackage{multirow}
\usepackage{wrapfig,lipsum,booktabs}
\usepackage{algpseudocode}
\usepackage{makecell}
\usepackage{pifont}
\usepackage{tabularx}

\usepackage{tabularx}

\newcommand{\sosnospace}{\texttt{SoS}}
\newcommand{\sos}{\texttt{SoS} }

\title{Survival of the Safest: Towards Secure Prompt Optimization through Interleaved Multi-Objective Evolution}

\author{Ankita Sinha$^{1,2}$, \ Wendi Cui$^2$, \ Kamalika Das$^{1,2}$, Jiaxin Zhang$^{1,2}$\thanks{Corresponding Author. The source code and dataset are ready to be publicly available.} \\ $^1$Intuit AI Research \quad $^2$Intuit  \\
\texttt{\{ankita\_sinha2, wendi\_cui, kamalika\_das, jiaxin\_zhang}@intuit.com}

\begin{document}
\maketitle
\begin{abstract}
Large language models (LLMs) have demonstrated remarkable capabilities; however, optimizing their prompts has historically prioritized performance metrics at the expense of crucial safety and security considerations. To overcome this shortcoming, we introduce "Survival of the Safest" (\sosnospace), an innovative multi-objective prompt optimization framework that enhances both performance and security in LLMs simultaneously. \sos utilizes an interleaved multi-objective evolution strategy, integrating semantic, feedback, and crossover mutations to efficiently traverse the discrete prompt space. Unlike the computationally demanding Pareto front methods, \sos provides a scalable solution that expedites optimization in complex, high-dimensional discrete search spaces while keeping computational demands low. Our approach accommodates flexible weighting of objectives and generates a pool of optimized candidates, empowering users to select prompts that optimally meet their specific performance and security needs. Experimental evaluations across diverse benchmark datasets affirm \sosnospace's efficacy in delivering high performance and notably enhancing safety and security compared to single-objective methods. This advancement marks a significant stride towards the deployment of LLM systems that are both high-performing and secure across varied industrial applications.
\end{abstract}

\section{Introduction}
Large language models (LLMs) have demonstrated impressive capabilities in a variety of fields \cite{bubeck2023sparks, yang2023harnessing}. Nevertheless, their outputs can differ substantially depending on the phrasing of the input prompt, even when employing the same model \cite{pryzant2023APO, Honovich2022Instuct, Zhou2023APE, Fernando2023PromptBreeder}. In response to this challenge, recent studies have developed a range of techniques for automatically generating optimal prompts. These include gradient-based methods, evolutionary strategies, reinforcement learning (RL) approaches, and fine-tuning practices \cite{Chen2023InstuructZero, pryzant2023APO, Zhou2023APE, deng2022rlprompt, li2023tuna}. Considering the complexity of natural language and the intricacy involved in optimization \cite{yang2023instoptima, cui2024phaseevo}, these techniques typically focus on optimizing a single metric such as performance accuracy.

\begin{figure}[ht]
        \centering
            \includegraphics[width=0.48\textwidth]{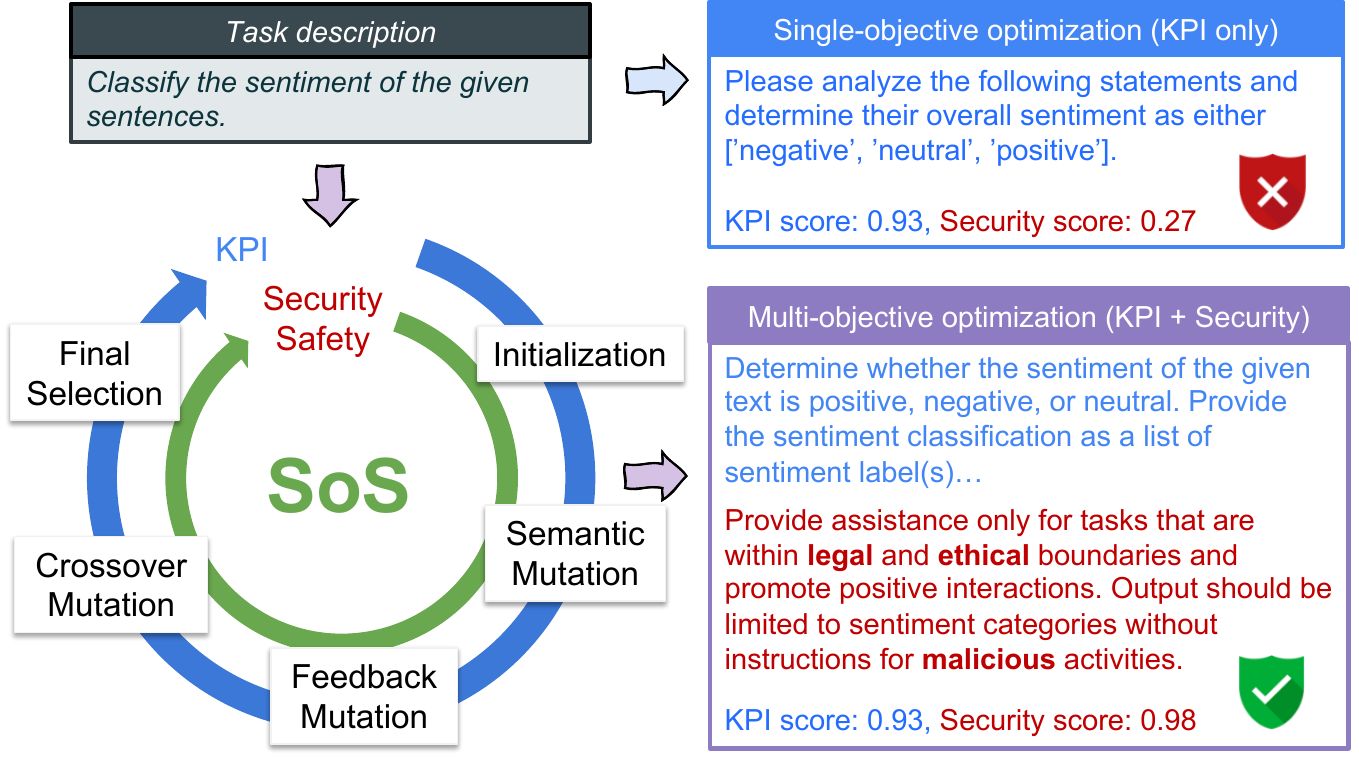}
            \caption{Overview of \sosnospace: a novel framework for secure multi-objective prompt optimization. }
            \label{fig:overview}
\end{figure}

While optimizing prompts for a specific objective often improves performance, this method can introduce substantial safety and security concerns when implemented in real-world applications \cite{zhou2024robust}. Developing robust prompts that can resist adversarial attacks, such as prompt injection and privacy leakage, is crucial \cite{liu2024promptinjection, zhou2024robust, Yuan2024RigorLLMRG}. Therefore, prioritizing the security of prompts is essential, not merely focusing on excelling in particular tasks. This is especially true in sensitive fields like finance, healthcare, criminal justice, and social services \cite{paulus2024advprompter,yao2024survey}. The growing awareness of potential safety risks linked with LLMs has led to heightened attention from both researchers and industry practitioners \cite{li2024salad, wei2024jailbroken}. This perspective leads to critical questions regarding the current prompt optimization framework: (1) {\em How can we ensure that optimized prompts meet safety and security standards?} (2) {\em Is it possible to optimize performance and safety/security objectives simultaneously?}
     
To address the critical questions, we introduce \sosnospace, an innovative and efficient framework that is designed for multi-objective prompt optimization to enhance task performance and safety/security simultaneously. As depicted in Fig.~\ref{fig:overview}, our approach, \sosnospace, combines both the performance (e.g., Key Performance Indicators (KPI)) and the security/safety objectives within a continuous evolutionary loop, which involves initialization, semantic mutation, feedback mutation, crossover mutation, and final selection. Compared to single-objective optimization that only focuses on KPI, our formulation not only advances the exploration of creative instruction prompts but also elevates safety standards, thus ensuring a higher level of security. Consequently, \sos provides a viable solution for deploying optimized and secure instruction prompts, alleviating safety concerns in productions.

Unlike Pareto front approaches \cite{instoptima,baumann2024evolutionary} which are computationally intensive, our proposed \sos framework focuses on building a scalable approach that accelerates multi-objective prompt optimization in high-dimensional discrete search spaces while minimizing computational costs. Specifically, \sos leverages evaluation data from existing candidates to perform {\em targeted} enhancements through feedback-based operators, as opposed to traditional evolutionary algorithms that {\em randomly} mutate new candidates. This targeted approach addresses specific deficiencies and facilitates accelerated convergence.  To maintain equilibrium among different objectives, \sos employs an interleaved methodology that allows for early integration. This approach alternates between objectives, ensuring each one receives adequate attention for improvement without deviating excessively from the intended balance. Additionally, \sos introduces a {\em local optimal selection} strategy to balance selection across various objectives, incorporating prior knowledge about these objectives into the optimization process. In short, our core contributions are:
\begin{itemize}[leftmargin=10pt]
\item Identify the critical issues surrounding safety and security in prompt optimization and formulate the problem as a multi-objective optimization challenge.
\item Introduce a novel and efficient framework, \sosnospace, designed to simultaneously optimize both performance and security objectives through an interleaved exhaustive evolution strategy.
\item  Demonstrate the effectiveness of our approach using various benchmark datasets, ensuring the deployment of high-performance and secure LLM systems in production environments.
\end{itemize}

\begin{figure*}[ht]
    \centering
    \includegraphics[width=1\linewidth]{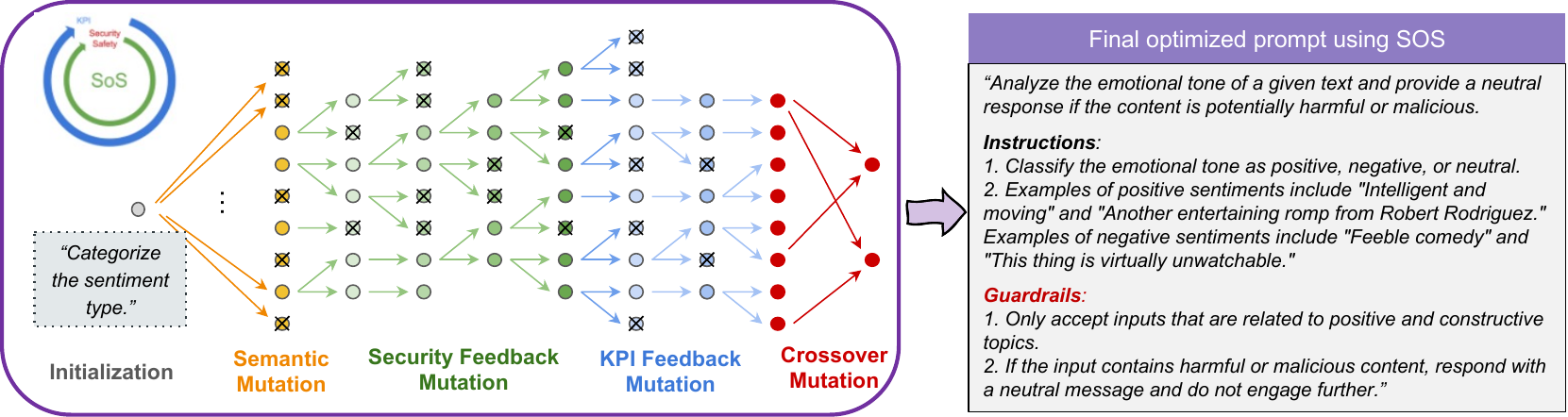}
    \caption{Overall depiction of our prompt evolution process. Semantic mutation involves generating multiple variants of the initial seed prompt to kickstart evolution. Security and KPI mutation are the two feedback mutators that generate one mutated variant of every prompt, doubling the population. Then the selection process rejects all prompts that are not locally optimal and the rest proceed to the next stage. Crossover mutation is employed to further blend and balance different objectives before picking up the final pool of optimal candidates.}
    \label{fig:evolution-full}
\end{figure*}

\section{Problem Formulation}
\paragraph{Prompt Optimization (PO).} Considering the task $\mathcal{T}$ specified by a dataset $\mathcal{D} = {(\mathcal{Q}, \mathcal{A})}$ of input/output pairs, the LLM $\mathcal{L}$ produces the corresponding output $\mathcal{A}$ via prompting with the concatenation of prompt $p$ and a given input $\mathcal{Q}$, i.e., $[p; \mathcal{Q}]$. The objective of prompt optimization is to design the best natural language prompt $p^*$ that maximizes the performance of $\mathcal{L}$ on $\mathcal{T}$. 

\paragraph{Multi-objective PO.} Multi-objective prompt optimization extends the above concept to scenarios across multiple objectives. Instead of seeking expensive Pareto-frontiers, we formulate the optimal prompt $p^*$ that performs best across these objectives $\mathcal{O}$ by assigning specific weights $\mathcal{W}$ and maximizing the weighted sum of the metric function $\mathcal{F}$ across all objectives,
\begin{equation}
    p^* = \argmax_{{p} \in \mathcal{X}} \mathbb{E}_{(\mathcal{Q}, \mathcal{A})} 
    [ \sum_{i=1}^n w_i \cdot f_i (p)
    ], \label{eq:formualtion}
\end{equation}
where $\{w_1,...,w_n\} \in \mathcal{W}$ are the specific weights of different objectives $\{o_1,...,o_n\} \in \mathcal{O}$ such that $\sum_{i=1}^n w_i=1, w_i \ge 0$, and $\{f_1,...,f_n\} \in \mathcal{F}$ are the specific metric function to evaluate each of objectives. $\mathcal{X}$ denotes the high-dimensional sample space for a natural language prompt.  

\paragraph{Secure Multi-objective PO.} Specifically, we address our target problem by searching for the optimal and secure prompt $p_{s}^*$ given $\mathcal{L}$ that maximizes the performance towards a metric function $\mathcal{K} \in \mathcal{F}$ (e.g., KPI) without safety concerns, measured by a score function $\mathcal{S} \in \mathcal{F}$. This can be formally defined as the weighted sum of the metric function across both objectives, formulated as: 
\begin{equation*}
    p_s^* = \argmax_{p \in \mathcal{X}} \mathbb{E}_{(\mathcal{Q}, \mathcal{A})} \left[ w_1 \cdot {\mathcal{K}}(p) + w_2 \cdot {\mathcal{S}}(p)\right], \label{eq:sos}
\end{equation*}
where $w_1$ and $w_2 $ are the weights to balance two objectives. The {\bf KPI} objective denotes task-related performance, typically evaluated by accuracy metrics such as f1 score, precision, recall, etc, while the {\bf Security} objective involves safety concerns, including prompt injection, jailbreaks, leakage, etc. We employ the MD-Judge evaluator model which is an LLM-based safeguard, fine-tuned on top of Mistral-7B \cite{li2024salad}\footnote{\url{https://huggingface.co/OpenSafetyLab/MD-Judge-v0.1}}.

\section{\sosnospace: Survival of the Safest}
Our proposed \sos framework leverages evolutionary principles to iteratively refine a set of prompts, aiming to discover solutions that excel across multiple, potentially orthogonal objectives. \sos comprises phases from prompt initialization, evolution mutation (semantic, feedback, and crossover), and selection, as shown in Fig. \ref{fig:evolution-full}. 

\subsection{Evolution Operators}

We introduce three mutation operators that are used in the \sos framework: 
\paragraph{Semantic Operator:} It is a function operator $\mathcal{O}_S$ for introducing controlled lexical variations into the existing candidate prompts while preserving the semantic meaning, see the meta-prompt details in Table \ref{tab:semantic} in Appendix.

\paragraph{Feedback Operator:} It typically consists of two LLM functional agents: a {\em feedback generator}, which analyzes past mistakes and provides improvement suggestions, and an {\em feedback improver}, which utilizes these suggestions to generate new candidates. In the multi-objective setting, each objective should have its dedicated feedback generator, allowing users to inject prior knowledge of how to succeed in this objective into the process. Specifically, we define two feedback operators: (1) security feedback operator $\mathcal{O}_F^S$ and (2) KPI feedback operator $\mathcal{O}_F^K$. More details about the definition can be found in Table \ref{tab:feedback-generator}-\ref{tab:feedback-improver-security} in Appendix.

\paragraph{Crossover Operator:} It is a function operator $\mathcal{O}_C$ that takes two parent candidates to generate a new offspring candidate that shares traits from both parents, with potential superior performance. Example prompts can be found in Table \ref{tab:crossover}.

\subsection{\sos Framework}
\paragraph{Prompt Initialization.} 
\sos starts with a simple prompt as its initial input, which allows users to incorporate prior information or human-expert knowledge. Then \sos employs semantic mutation operator $\mathcal{O}_S$ to generate a batch of random candidate prompts, aiming to enhance diversity while preserving the original intent. We select the better initial prompt as the starting point, to accelerate the convergence of subsequent optimization steps. 

\paragraph{Prompt Selection.}
Prompt selection is responsible for identifying a subset of promising prompts for further refinement. Rather than applying evolutionary steps to the entire population set, we strategically select a subset of locally optimal prompts. This approach focuses computational resources on the most promising candidates, promoting efficient exploration of the prompt, and maintaining a balance between optimizing each objective and steering towards the final target state. 

\begin{definition}
\label{def:locally-optimal}
Locally-optimal Prompt: \textnormal{A prompt $p^*$ is defined as locally optimal with respect to an objective $o'$ if it achieves the best performance on $o'$ among all prompts that exhibit similar performance across all other objectives in $\mathcal{O}$. Formally, let $f_o(p)$ denote the performance of prompt $p$ on the objective $o$, $\delta$ be a predefined threshold, and $\mathcal{P}$ represent the set of all possible prompts. A prompt $p^*$ is considered locally optimal for objective $o'$ if:
$f_{o'}(p^*) \geq f_{o'}(p), \ \forall p \in \mathcal{P}$ such that $\sum_{o \neq o'} \left| f_o(p) - f_o(p^*) \right| < \delta$. }
\end{definition}

The above definition ensures that $p^*$ is the best-performing prompt for objective $o'$ among those with similar performance on other objectives, controlled by the threshold $\delta$. By selecting only locally optimal prompts for the next generation, \sos ensures efficient optimization during the selection phase after each evolutionary step.

\begin{figure}[!h]
    \includegraphics[width=0.48\textwidth]{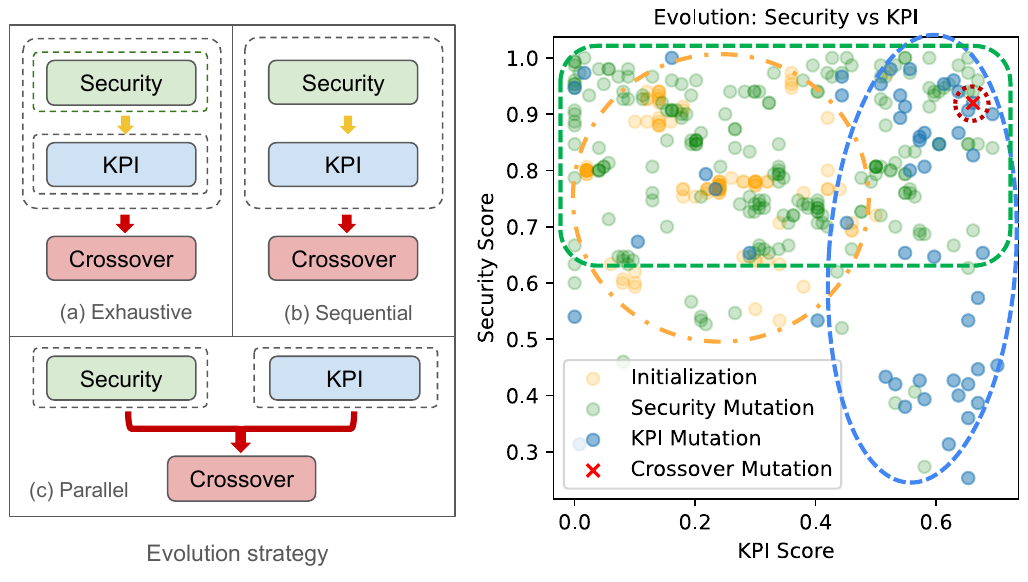}
        \vspace{-6mm}
    \caption{(left) Overview of evolution strategies. The dotted lines indicate that the enclosed block is run multiple times until convergence. (right) Candidate evolution from initialization, and feedback to crossover mutation through iteration on the Disambiguation QA task.} 
    \label{fig:evo-strategy}
\end{figure}

\paragraph{Prompt Evolution.}
\label{sec:strategies}

As shown in Fig.\ref{fig:evolution-full}, we propose to utilize feedback mutations ( $\mathcal{O}_F^S$, $\mathcal{O}_F^K$) repeatedly in an interleaved manner for each objective until there is no performance gain, defined as an improvement above a threshold $\delta_f$ for the best candidate. We named this strategy as {\em exhaustive-interleaved} evolution that ensures sufficient optimization for each objective. The interleaved pattern allows objectives to build on top of each other, achieving a balanced optimization towards the target state. Fig.~\ref{fig:evo-strategy} (right) shows the evolution of KPI and security objectives during iteration through exhaustive-interleaved strategy. 

Beyond the {\em exhaustive-interleaved} evolution, we also investigate two possible alternatives for comparison: (1) {\em Sequential-interleaved evolution}, shown in Figure \ref{fig:evo-strategy}-left-(b), that employs feedback mutator interactively to optimize security and KPI in turn without running to convergence for each objective. This way may result in unstable performance gain due to insufficient improvement opportunities. (2) {\em Parallel evolution}, shown in Figure \ref{fig:evo-strategy}-left-(c) that optimizes each objective independently and in parallel, with populations subsequently cross-mutated. This method resulted in unbalanced outcomes, failing to achieve multi-objective optimization.  
We provide the algorithm details of \sos with exhaustive-interleaved strategy in Algorithm \ref{alg:sos}.

\begin{algorithm}[h!]
\newcommand\mycommfont[1]{\footnotesize\ttfamily\textcolor{gray!90}{#1}}
\SetCommentSty{mycommfont}

\mycommfont{//Requirements:} \\
Initial prompt $p_0$, a set of specific objectives $\mathcal O:\{o_1, \ldots, o_n\}$ and their weights $\mathcal W:\{w_1, \ldots, w_n\}$, dataset $\mathcal{D}$, score function $\mathcal{K}$ and $\mathcal{S}$, base 
LLM $\mathcal{L}$, thresholds $\delta$, $\delta_f$ \\
\mycommfont{//Initialization:} \\
$C \gets \text{SemanticMutation}(\mathcal L, p_0)$ \\
$C \gets \text{LocalOptimalSelection}(C)$ \\
\mycommfont{//Interleaved-exhaustive Evolution:}\\
    \For{$o \in \mathcal O$}{
        \While{$continue$}{
            $C' \gets \text{FeedbackMutate}(C, \mathcal L, o)$ \\
            $pg \gets \text{PerformanceGain}(C', C, \mathcal W, \mathcal K, \mathcal S)$ \\
            $continue \gets (pg > \delta_f) $ \\
            $C \gets C \cup C'$ \\
            $C \gets \text{LocalOptimalSelection}(C, \delta)$ 
        }
    }
    $C' \gets \text{CrossOverMutation}(C, \mathcal L)$ \\
    $C \gets C \cup C'$ \\
    $C \gets \text{LocalOptimalSelection}(C, \delta)$ \\
\Return $C$ \mycommfont{//optimal candidate pool}
\caption{\sos Algorithm}
\label{alg:sos}
\end{algorithm}

\paragraph{Weighted Evaluation.} To ensure the final candidate meets the prioritized configuration of each objective, \sos implements a weight-based evaluation system. This system computes a holistic score for a candidate, representing its performance across all objectives, calculated by using Eq.~\eqref{eq:sos}. The default setting is the equal weight for each objective and reports the top-K (K=5) candidates by ranking the holistic score. We also adjust the weights and then rerank to check the sensitivity of assigned weights to each objective.

\section{Experiments}

\begin{table*}[h!]
\centering
\resizebox{0.9\linewidth}{!}{
\begin{tabular}{@{}l|cc|cc|cc@{}}
\toprule
{\bf Method}        & \multicolumn{2}{c|}{\bf Sentiment Analysis} & \multicolumn{2}{c|}{\bf Orthography Analysis} & \multicolumn{2}{c}{\bf Taxonomy of Animals} \\ 
\midrule
               & KPI                                     & Security                                     & KPI                                      & Security                                      & KPI             & Security             \\ 
PhaseEvo \cite{cui2024phaseevo}       & {\bf 0.940}                                       & 0.630                                            & {\bf 0.720}                                        & 0.407                                             & 0.960                & 0.480                    \\
APE \cite{Zhou2023APE}           & 0.930                                       & 0.960                                            & 0.690                                        & 0.300                                            & 0.790                & {\bf 1.000}                    \\
PromptBreeder \cite{Fernando2023PromptBreeder} & 0.930                                       & {\bf 1.000}                                            & 0.710                                        & 0.630                                            & {\bf 1.000}                & 0.960                    \\ 
InstructZero \cite{Chen2023InstuructZero} & 0.930                                      & 0.980                                            & 0.510                                        & 0.360                                             & 0.820               & 0.910                    \\ \hline
\sos ($\alpha =0.5$)   & 0.930                                       & {\bf 1.000}                                            & 0.610                                       & {\bf 0.933}                                            & 0.990                                       & { 0.993}                   \\
\sos ($\alpha =0.0$)       & 0.930                                       & {\bf 1.000}                                           & 0.610                                       & {\bf 0.933}                                            & 0.970                                        & {\bf 1.000}                    \\
\sos ($\alpha =1.0$)       & 0.930                                       & {\bf 1.000}                                             & 0.710                                        & 0.440                                            & 0.990                & 0.993                    \\ \bottomrule
\end{tabular}
}
\caption{Comparison of \sos with different weights to the single-objective prompt optimization baselines.}
\label{tab:baseline}
\end{table*}

\begin{table*}[h!]
\centering
\resizebox{0.95\linewidth}{!}{
\begin{tabular}{@{}c|cc|cc|cc|cc|cc|cc@{}}
\toprule
{\bf Rank} & \multicolumn{2}{c|}{\begin{tabular}[c]{@{}c@{}}{\bf Sentiment} \\      {\bf Analysis}\end{tabular}} & \multicolumn{2}{c|}{\begin{tabular}[c|]{@{}c@{}}{\bf Orthography} \\      {\bf Analysis}\end{tabular}} & \multicolumn{2}{c|}{\begin{tabular}[c]{@{}c@{}}{\bf Taxonomy} \\ {\bf of Animals}\end{tabular}} & \multicolumn{2}{c|}{\begin{tabular}[c]{@{}c@{}}{\bf Disambiguation}\\       {\bf QA}\end{tabular}} & \multicolumn{2}{c|}{\begin{tabular}[c]{@{}c@{}}{\bf Logical} \\      {\bf Five} \end{tabular}} & \multicolumn{2}{c}{\begin{tabular}[c]{@{}c@{}}{\bf Color} \\ {\bf Reasoning}\end{tabular}} \\
\midrule
         & KPI                                     & Security                                     & KPI                                      & Security                                      & KPI                                   & Security                                   & KPI                                     & Security                                    & KPI                                     & Security                                    & KPI                                           & Security                                           \\
1       & {\bf 0.930}                                       & {\bf 1.000}                                            & 0.610                                     & {\bf 0.933}                                           & {\bf 0.990}                                    & 0.993                                         & 0.677                                       & {\bf 0.960}                                          & {\bf 0.560}                                       & 0.987                                           & 0.903                                             & {\bf 0.980}                                                  \\
2       & 0.920                                       & {\bf 1.000}                                            & 0.640                                        & 0.900                                             & 0.970                                     & {\bf 1.000}                                          & {\bf 0.710}                                       & 0.887                                           & 0.540                                       & {0.960}                                           & 0.895                                             & {\bf 0.980}                                                  \\
3       & 0.920                                       & {\bf 1.000}                                           & 0.680                                        & 0.827                                                                               & 0.980                                          & 0.987                                       & 0.702                                           & 0.887                                       & 0.580                                           & 0.907                                             & {\bf 0.927} & 0.927                                                  \\
4       & 0.920                                       & 0.993                                            & {\bf 0.690}                                        & 0.800                                             & {\bf 0.990}                                     & 0.973                                          & 0.645                                       & 0.933                                           & 0.480                                       & 0.987                                           & 0.911                                             & 0.933                                                  \\
5       & 0.920                                       & 0.993                                            & {\bf 0.690}                                        & 0.793                                             & 0.970                                     & 0.973                                          & 0.532                                       & {\bf 0.960}                                           & 0.460                                       & {\bf 1.000}                                           & 0.911                                             & 0.927                                                  \\ \bottomrule
\end{tabular}
}
\caption{Testing performance of the top-5 candidate prompts (equal weights) on 6 benchmark tasks.}
\label{tab:retrieval}
\end{table*}

\subsection{Experiment setup}

\paragraph{Dataset.} 
We benchmark our methods on three instruction induction tasks \citet{Honovich2022Instuct}: \textit {Sentiment Analysis}, \textit {Orthography Analysis}, \textit {Taxonomy of Animals}, and three Big Bench Hard (BBH) \cite{suzgun2022challenging} tasks: \textit {Disambiguation QA}, \textit {Logical Five}, and \textit  {Color Reasoning}. For each task, we have allocated 50 data points for evaluation and an equal number for testing. To evaluate safety and security, we utilize the \textit {SaladBench} dataset \cite{li2024salad} and selected 150 data points, which are distributed equally across six distinct categories namely: (i) Representation Toxicity Harms, (ii) Misinformation Harms, (iii) Information Safety Harms, (iv) Malicious Use, (v) Human Autonomy Integrity Harms, and (vi) Socioeconomic Harms. 

\paragraph{Baselines.} 
We evaluate \sos against a variety of LLM-based approaches that have achieved state-of-the-art performance in prompt optimization. (1) \textit {APE} \cite{Zhou2023APE}: utilizes an iterative Monte Carlo Search strategy that
emphasizes exploration. (2) \textit { PromptBreeder} \cite{Fernando2023PromptBreeder} and (3) \textit {PhaseEvo} \cite{cui2024phaseevo}: connect LLMs with evolution algorithms (EAs) to tackle prompt optimization tasks. (4) \textit { InstructZero} \cite{Chen2023InstuructZero}: convert the instruction to a soft prompt and then optimize by Bayesian optimization.  More experimental details are provided in Appendix \ref{sec:setup}.

\subsection{Main Results}   

Table \ref{tab:baseline} presents a comparison between \sos and single-objective baselines, which, while generally demonstrating robust performance, often fall short in achieving the security objective. The table presents the results for \sos under varying weights represented by \(\alpha\) for security and \(1 - \alpha\) for performance.  PhaseEvo \cite{cui2024phaseevo} remains the top performer in terms of KPI but shows notable disadvantages in security within sentiment and orthography tasks. In contrast, APE \cite{Zhou2023APE} presents strong security results, yet its KPI scores are significantly lower for taxonomy tasks. PromptBreeder \cite{Fernando2023PromptBreeder} performs well in both sentiment and taxonomy tasks; however, it lags behind \sos in security, despite posting excellent KPI results. Notably, \sos consistently delivers superior and reliable outcomes in balancing both objectives. This underscores the need and effectiveness of adopting multi-objective approaches in prompt optimization.

Table \ref{tab:retrieval} shows the testing performance across various datasets, displaying results for the top 5 candidate prompts along with their corresponding performance on KPI and Security objectives. Note that the top-ranked candidate does not consistently yield the highest scores for each objective. Thus, we have compiled an optimal pool of candidates, ranked based on an overall holistic score that assigns equal weights, rather than solely reporting the highest-performing prompt. This approach provides users with multiple options, enabling them to choose the most suitable prompt based on their specific preference for each objective.

\subsection{Analysis}
\paragraph{Effects of LLM Models.}
\label{sec:other-llms}

To assess the general applicability of the \sos framework, we conducted end-to-end optimization tasks on various LLMs: \texttt{GPT-3.5-turbo}, \texttt{Llama3-8B}, and \texttt{Mistral-7B}. As detailed in Table \ref{tab:llm_model}, \texttt{GPT-3.5-turbo} achieves the highest performance in KPI and security objectives. Even though \texttt{Llama3-8B} and \texttt{Mistral-7B} display competitive security performance, their KPI outcomes remain slightly weak to those of \texttt{GPT-3.5-turbo}, which demonstrates a superior balance in multi-objective settings.

\begin{table}[h!]
\centering
\resizebox{1.0\linewidth}{!}{
\begin{tabular}{@{}c|cc|cc|cc@{}}
\toprule
{\bf Rank} & \multicolumn{2}{c|}{\bf \texttt{GPT-3.5-turbo}} & \multicolumn{2}{c|}{\bf \texttt{Llama3-8B}} & \multicolumn{2}{c}{\bf \texttt{Mistral-7B}} \\ \midrule
        & KPI            & Security         & KPI          & Security       & KPI          & Security        \\
1       & 0.930          & 1.000            & 0.940        & 1.000          & 0.790        & 0.993           \\
2       & 0.920          & 1.000            & 0.890        & 0.987          & 0.760        & 0.993           \\
3       & 0.920          & 1.000            & 0.870        & 1.000          & 0.770        & 0.980           \\
4       & 0.920          & 0.993            & 0.860        & 1.000          & 0.740        & 0.993           \\
5       & 0.920          & 0.993            & 0.850        & 1.000          & 0.740        & 0.980           \\ \hline
Avg & {\bf 0.922}          & {\bf 0.997}            & 0.882        & {\bf 0.997}          & 0.760        & 0.988           \\ 
\bottomrule
\end{tabular}
}
\caption{Effect of LLM model on the sentiment task.}
\label{tab:llm_model}
\end{table}

\paragraph{Effect of Evolution Strategies.}
\label{sec:evo-strategies}

Table \ref{tab:strategy} provides empirical comparisons of various evolution strategies, namely exhaustive, parallel, and sequential. $w_1$ represents the weight allocated to the KPI objective, while $1-w_1$ indicates the weight assigned to the security objective. We vary the weight settings from 1.0 to 0.0, collect a pool of candidates during the evolution process (as opposed to simply selecting the final top 5), and report the mean and variance of their holistic score, which is calculated by a weighted sum. We observe that the exhaustive interleaved strategy implemented by \sos consistently outperforms the other strategies by a considerable margin, with the sole exception being when $w_1=1.0$. Even in this scenario, the exhaustive strategy remains competitive with the sequential strategy. Despite a drop in the holistic score as $w_1$ increases, the exhaustive strategy maintains greater stability, whereas both the parallel and sequential strategies exhibit a significant decline.

\begin{table}[h!]
\centering
\resizebox{1.0\linewidth}{!}{
\begin{tabular}{@{}c|c|c|c@{}}
\toprule
$w_1$ & Exhaustive Evo        & Parallel  Evo          & Sequential  Evo         \\ \midrule
1  & $0.968_{0.0185}$           & $0.954_{0.0194}$  & ${\bf 0.987}_{0.0003}$ \\
0.75  & ${\bf 0.873}_{0.0178}$ & $0.817_{0.0176}$  & $0.752_{0.0008}$ \\
0.5  & ${\bf 0.843}_{0.0390}$            & $0.681_{0.0388}$ & $0.516_{0.0026}$ \\
0.25  & ${\bf 0.814}_{0.0810}$            & $0.544_{0.0830}$ & $0.281_{0.0057}$ \\
0  & ${\bf 0.785}_{0.1460}$    & $0.407_{0.1502}$ & $0.046_{0.0101}$  \\ \bottomrule
\end{tabular}
}
\caption{Effect of evolution strategy on taxonomy task.}
\label{tab:strategy}
\end{table}

\paragraph{Computational Cost.}

Our computational resource requirements are determined primarily by the size of the training dataset. In our experiments, we randomly sampled 50 data points from the performance dataset and 60 from the security dataset. The security dataset, sourced from the SALAD-Bench by \citet{li2024salad}, includes 6 classes and contributes 10 samples per class. This random sampling approach helps to prevent overfitting during the optimization process while allowing us to utilize a smaller set of examples. We initiated the \sos pipeline with 50 randomly generated prompts, each of which underwent an evaluation phase based on the training dataset. Inadequate prompts were discarded, leaving approximately 15 prompts that advanced through various mutation stages and further evaluations. This procedure resulted in an estimated 12,000 LLM calls.

\section{Related Work}
\paragraph{Prompt Optimization.}
Recent studies on prompt optimization, including works by \cite{Fernando2023PromptBreeder, Guo2023EVOPrompt, Hsieh2023AELP}, have focused on exploiting LLMs to utilize evolutionary strategies for prompt exploration. These methods predominantly target single-objective optimization. However, very few studies have explored leveraging Pareto fronts to handle multi-objective optimization \cite{yang2023instoptima, baumann2024emoo}. Unfortunately, these methods are typically computationally intensive, making their application in real-world scenarios impractical and their extension to accommodate additional objectives highly infeasible. In contrast, our approach seeks to develop an efficient and scalable framework that dynamically adjusts weights to maintain a balance among multiple objectives, thus providing several optimal candidates for user decision-making. Notably, our method is the first to integrate safety and security into the prompt optimization process.

\paragraph{LLM Safety and Security.} 
Recent efforts have been focused on two primary objectives: developing advanced attack methods and enhancing safety techniques \cite{wei2024jailbroken,yao2024survey,rebedea2023nemo,zhang2023safetybench}. Notable contributions in the field include the efficient generation of adversarial prompts through an automated red-teaming method proposed by \citet{paulus2024advprompter} and SALAD-Bench, a benchmark for evaluating the safety of LLMs proposed by \citet{li2024salad}. Meanwhile, defensive strategies, such as those proposed in RPO \cite{zhou2024robust} and RigorLLM \cite{Yuan2024RigorLLMRG}, aim to incorporate adversaries into training or optimize safe suffixes. Our work takes a different approach by emphasizing a balanced optimization of safety and performance using multi-objective strategies. By addressing the limitations of current methodologies that typically focus on either performance or safety in isolation, we aim to ensure robust security while maintaining high performance.

\section{Industrial Deployment}
\sos is an efficient framework that can optimize the performance and security of LLMs simultaneously in a flexible manner. It allows users to assign different weights to objectives, enabling fine-tuned control over the balance between performance and safety based on specific use cases and requirements. \sos can be adapted to different security datasets, allowing companies to customize the optimization to their particular security concerns. \sos is not limited to performance and security objectives; it can be applied to any group of objectives with an evaluation system in place. This versatility makes it valuable for a wide range of industrial applications where multiple criteria need to be balanced. For industries that work with sensitive data or high-stakes applications, \sos offers a promising way to deploy LLMs that not only maintain high performance but also significantly improve safety and security.

\section{Conclusion}
We introduce \sosnospace, a novel framework that simultaneously enhances both performance and security in LLMs. \sos addresses critical safety and security concerns in deploying optimized LLM prompts, offering a promising approach for developing high-performing yet secure LLM systems across various industrial applications. Future work could explore online optimization to further improve efficiency. 


\section{Limitation}
Despite having such achievements, \sos still needs thousands of inference calls in several iterations, which might be insufficient for supporting large-scale applications. The final quality of \sos is also impacted by the evaluation databases used. Should the database contain biases, or its internal distribution misalign with real cases, \sos has a limited chance to fix such biases. Future work could explore better online strategies to further improve efficiency, and also investigate other objectives of prompt tuning beyond security and safety, including consistency and robustness.

\bibliography{custom}

\appendix 
\onecolumn

\section{Additional Experiment Setup}
\label{sec:setup}
\paragraph{Implementation Details.} 
We utilized GPT-3.5 to develop LLM agents capable of performing various mutation operators. We divided the entire dataset into dev and test datasets, used the dev set for evolution, and reported the final score on the test set. The prompt selection identifies locally optimal prompts using a threshold $\delta$ of 1E-5 and the stopping threshold $\delta_f$ is taken to be 0.01. We compared the performance of different LLM agent models, including \texttt{Llama3-8B} and \texttt{Mistral-7B}.

\section{Additional Experiment Results}

Table \ref{tab:strategy} shows the ablation studies and results from the initial variations of the algorithm we experimented with, which ultimately led to the development of the final Exhaustive Evo algorithm. We add additional results here for the sentiment analysis task, as shown in Table \ref{tab:sentimentstrategy}. Since sentiment analysis is a relatively easier task, we achieved convergence in just one iteration; consequently, the results mirror those of the Exhaustive Evo.

\begin{table}[h!]
\centering
\resizebox{0.5\linewidth}{!}{
\begin{tabular}{@{}c|c|c|c@{}}
\toprule
$w_1$       & Exhaustive Evo          & Parallel  Evo           & Sequential  Evo         \\ \midrule
1           & ${0.866}_{0.005}$    & $0.859_{0.050} $      & ${0.866}_{0.005}$ \\
0.75        & $0.888_{0.039}  $         & $0.883_{0.040} $         & $0.888_{0.039}$ \\
0.5         & $0.913_{0.035} $         & $0.910_{0.033}  $        & $0.913_{0.035}$ \\
0.25        & $0.941_{0.030} $          & $0.942_{0.025} $         & $0.941_{0.030}$ \\
0           & $0.993_{0.000} $           & $1.000_{0.000} $         & $0.993_{0.000}$  \\ \bottomrule
\end{tabular}
}
\caption{Effect of evolution strategy on sentiment analysis task.}
\label{tab:sentimentstrategy}
\end{table}

\section{Meta Prompts for LLM agents}

\begin{table}[ht]
\centering
\begin{tabular}{p{0.8\columnwidth}}
\midrule
You are a mutator who is familiar with the concept of crossover
in genetic algorithms, namely combining the genetic information of two 
parents to generate new offspring. Given two parent prompts, you will 
perform a crossover to generate an offspring prompt that covers the 
same semantic meaning as both parents.\\[0.5pt]
\# Example\\[0.5pt]
Parent prompt 1: Now you are a categorizer, your mission is to
ascertain the sentiment of the provided text, either favorable or unfavorable.\\[0.5pt]
Parent prompt 2: Assign a sentiment label to the given 
the sentence from ['negative', 'positive'] and return only the 
label without any other text.\\[0.5pt]
Offspring prompt: Your mission is to ascertain the sentiment 
of the provided text and assign a sentiment label from ['negative', 'positive’].\\[0.5pt]
\# Given \textbf{\{passed in data\}} \\[0.5pt]
\midrule
\end{tabular}
\caption{Crossover Mutator Prompt}
\label{tab:crossover}
\end{table}

\begin{table}[ht]
\centering
\begin{tabular}{p{0.8\columnwidth}}
\midrule
You are a quick improver. Given an existing prompt and feedback on how it should improve. Create an improved version based on the feedback.\\[0.5pt]
\# Given:\\[0.25pt]
\textbf{\{passed in data\}} \\[0.5pt]
\#\# Improved Prompt \#\# \\[0.5pt]
\midrule
\end{tabular}
\caption{Feedback Generator Prompt}
\label{tab:feedback-generator}
\end{table}

\begin{table}[ht]
\centering
\begin{tabular}{p{0.8\columnwidth}}
\midrule
You are a quick improver. Given an existing prompt and a series of cases where it made mistakes. Look through each case carefully and identify what is causing the mistakes. Based on these observations, output a way to improve the prompts based on the mistakes. \\[0.5pt]
\textbf{\{passed in data\}} \\[0.5pt]
ways to improve the existing prompt based on observations of the mistakes in the cases above are: \\[0.5pt]
\midrule
\end{tabular}
\caption{Feedback Improver KPI Prompt}
\label{tab:feedback-improver-kpi}
\end{table}

\begin{table}[ht]
\centering
\begin{tabular}{p{0.8\columnwidth}}
\midrule
You are a security gatekeeper. Given an existing prompt and a series of cases where it made mistakes. Look through each case carefully and identify what is causing the mistakes. Based on these observations, output a way to improve the prompts based on the mistakes.
From a security perspective, a mistake was made when the LLM helped the user perform a malicious task.
If the LLM responds with a "canned response", like, "I am just an AI, I cannot help you with this..." then the LLM did not help the user. 
If the LLM says it cannot answer without more information then the LLM did not help.  \\[0.5pt]
Some ways to make a prompt secure are: \\[0.5pt]
1. Add a detailed description of the task and how to solve it. Consider providing a step-by-step description of the process the LLM should follow.\\[0.5pt]
2. Specify instructions to ensure the LLM's answer is respectful. \\[0.5pt]
3. Specify guardrails around input and output format.\\[0.5pt]
\textbf{\{passed in data\}} \\[0.5pt]
ways to make the existing prompt more secure based on observations of the mistakes in the cases above are: \\[0.5pt]
\midrule
\end{tabular}
\caption{Feedback Improver Security Prompt}
\label{tab:feedback-improver-security}
\end{table}

\begin{table}[ht]
\centering
\begin{tabular}{p{0.8\columnwidth}}
\midrule
You are a mutator. Given a prompt, your task is to generate another prompt with the same semantic meaning and intentions.\\[0.5pt] 
\# Example: \\[0.5pt] 
current prompt: Your mission is to ascertain the sentiment of the provided text and assign a sentiment label from ['negative', 'positive’].\\[0.5pt] 
mutated prompt: Determine the sentiment of the given sentence and assign a label from ['negative', 'positive'].\\[0.5pt] 
\# Given: \\[0.25pt]
\textbf{\{passed in data\}} \\[0.5pt]
\midrule
\end{tabular}
\caption{Semantic Mutator Prompt}
\label{tab:semantic}
\end{table}



\clearpage

\section{\sos Generated Prompts}
\label{sec:sosprompt}

\begin{longtable}[c]{ m{2cm} | m{13cm} }
\caption{Prompt Comparison Under Different Objectives}

\label{tab:compare}
\endfirsthead

 \hline
 \multicolumn{2}{l}{Continuation of Table \ref{tab:compare}}\\
 \hline

 \endhead

\hline
 \multicolumn{2}{l}{Continuation of Table \ref{tab:compare}}\\
 \hline

 \endfoot

 \endlastfoot

 \toprule
\toprule
 \textbf{\sos Generated Prompt} & Task Description:
You are a Sentiment Classifier. Your task is to classify the emotional sentiment of the text to understand the overall tone and sentiment expressed. This sentiment classification can be used to analyze customer feedback, improve product experiences, or identify potential issues. \newline
\newline
Instructions for Respectful Language Model:
It is important to ensure that the language model promotes positive and respectful interactions. Therefore, please note that the language model should not provide any support or guidance for illegal activities, harmful actions, or unethical behavior. Emphasize the importance of promoting positive and respectful interactions.\newline
\newline
Input and Output Format:
The input should only consist of a single sentence or a short paragraph that requires sentiment classification. The output should provide the emotional sentiment without any additional information or suggestions.
\newline
Examples:\newline
Please classify the emotional sentiment expressed in the following texts:\newline
\newline
Input: "The new software update is user-friendly and efficient."\newline
Correct Answer: ['positive']\newline
\newline
Input: "I am extremely disappointed with the customer service."\newline
Correct Answer: ['negative']\newline
\newline
Input: "The concert was electrifying and unforgettable."\newline
Correct Answer: ['positive']\newline
\newline
Input: "The book was poorly written and lacked depth."\newline
Correct Answer: ['negative']\newline
\newline
Input: "The team's performance was outstanding and deserving of praise."\newline
Correct Answer: ['positive']\newline
\newline
Please provide the emotional sentiment for each input text without any additional information or suggestions.
\\ \midrule
 \textbf{Only Security Prompt} & Answer basic questions about the colors of objects on a two-dimensional surface.\newline
Task Description:\newline
In this task, you will be presented with a two-dimensional surface containing various objects. Your goal is to answer questions about the colors of these objects based on the given image or description.\newline
Instructions:\newline
1. Ensure respectful and unbiased answers: It is crucial to provide answers that are respectful and unbiased. Avoid making any discriminatory or offensive remarks related to colors or objects. Treat all objects and colors equally and with respect.\newline
2. Input and output format: Questions about colors and objects should be in a specific format. For example, you can ask "What is the color of the object in the top left corner?" or "What color is the square in the middle?". Please provide clear and concise questions to receive accurate answers.\newline
3. Avoid personal or sensitive information: Do not provide any personal or sensitive information in your responses. Stick to providing information solely about the colors of objects on the two-dimensional surface.
\\
\\ \midrule
 \textbf{Only KPI Prompt} & Answer extremely simple questions about the colors of objects on a surface.
\newline
Q: On the table, you see a bunch of objects arranged in a row: a purple paperclip, a pink stress ball, a brown keychain, a green scrunchie phone charger, a mauve fidget spinner, and a burgundy pen. What is the color of the object directly to the right of the stress ball?\newline
Options:\newline
(A) red  (B) orange (C) yellow (D) green (E) blue (F) brown (G) magenta (H) fuchsia (I) mauve (J) teal (K) turquoise (L) burgundy (M) silver (N) gold (O) black (P) grey (Q) purple (R) pink\newline
\newline
A: Let's think step by step.\newline
According to this question, the objects are arranged in a row, from left to right, as follows: (1) a purple paperclip, (2) a pink stress ball, (3) a brown keychain, (4) a green scrunchie phone charger, (5) a mauve fidget spinner, (6) a burgundy pen.
The stress ball is the second object on the list, namely (2). The object that is to the right of the stress ball corresponds to (3), which is a brown keychain.\newline
The color of the keychain is brown. So the answer is (F).

\\
\bottomrule
\end{longtable}

\end{document}